# All-sky Radio SETI

**M.A. Garrett**[*1-3]
[1]*Jodrell Bank Centre for Astrophysics, School of Physics & Astronomy, The University of Manchester, Alan Turing Building, Oxford Road, Manchester, M13 9PL, UK.*
[2]*Leiden Observatory, Leiden University, Post box 9513, 2300RA Leiden, The Netherlands.*
[3]*ASTRON, Netherlands Institute for Radio Astronomy, Post box 2, 7990AA, Dwingeloo, The Netherlands.*
E-mail: Michael.Garrett@manchester.ac.uk

**A. Siemion**[1-3]
[1]*Dept. of Astronomy & Radio Astronomy Lab, Univ. of California, Berkeley, USA.*
[2]*ASTRON, Netherlands Institute for Radio Astronomy, Post box 2, 7990AA, Dwingeloo, The Netherlands.*
[3]*Dept. Of Astrophysics IMAPP, Radboud University, PO Box 9010, 6500GL Nijmegen, The Netherlands.*
E-mail: siemion@berkeley.edu

**W.A. van Cappellen**
*ASTRON, Netherlands Institute for Radio Astronomy, Postbus 2, 7990AA, Dwingeloo, The Netherlands.*
E-mail: cappellen@astron.nl

Over the last decade, Aperture Arrays (AA) have successfully replaced parabolic dishes as the technology of choice at low radio frequencies – good examples are the MWA, LWA and LOFAR. Aperture Array based telescopes present several advantages, including sensitivity to the sky over a very wide field-of-view. As digital and data processing systems continue to advance, an all-sky capability is set to emerge, even at GHz frequencies. We argue that assuming SETI events are both rare and transitory in nature, an instrument with a large field-of-view, operating around the so-called "water-hole" (~ 1-2 GHz), might offer several advantages over contemporary searches. Sir Arthur C. Clarke was the first to recognise the potential importance of an "all-sky" radio SETI capability, as presented in his book, Imperial Earth. As part of the global SKA (Square Kilometre Array) project, a Mid-Frequency Aperture Array (MFAA) prototype known as MANTIS (Mid-Frequency Aperture Array Transient and Intensity-Mapping System) is now being considered as a precursor for SKA-2. MANTIS can be seen as a first step towards an "all-sky" radio SETI capability at GHz frequencies. This development has the potential to transform the field of SETI research, in addition to several other scientific programmes.

---

[*] Speaker

1.     **SETI and Aperture Arrays**

The Search for Extraterrestrial Intelligence (SETI) is a scientific pursuit that has been undergoing a significant revival for several years now, accelerated by initiatives such as the recent establishment of the Breakthrough Listen (BL) project [1]. BL is expected to invest ~ 100M$ in the field over the next 10 years, and has recently embarked on ambitious surveys of the sky using the Green Bank and Parkes radio telescopes (see Siemion et al. these proceedings). In addition, the rapid development of radio astronomy capabilities in general, in key areas such as sensitivity, wavelength coverage, and spatial/spectral/temporal resolution is also very encouraging. It seems humankind is now better placed than ever before to make a SETI detection using these new instruments.

The BL project currently focuses on surveys using large parabolic telescopes. While this maximises sensitivity in a particular direction, the field-of-view is set by the diameter of the dish – so bigger, more sensitive telescopes, produce narrower beams. The use of aperture array technology promises to create telescopes with both good sensitivity and a wide field-of-view. Aperture arrays are formed from a network of many thousands of individual antennas that are combined together electronically, with appropriate delays applied to each antenna, maximising sensitivity in a particular direction on the sky. Since each individual antenna is sensitive to the whole sky (or something close to that), additional electronics can form multiple-beams, resulting in sensitivity (gain) in many different directions simultaneously. In principle, an aperture array can provide all-sky coverage, given the right architecture and sufficient electronic processing power (data sampling, digitisation and beam forming). The concept of forming a parabolic dish electronically permits us to consider an aperture of gigantic proportions, pointing in many directions simultaneously. Such a huge and highly sensitive aperture also has no moving parts and can point to any area of sky almost instantaneously.

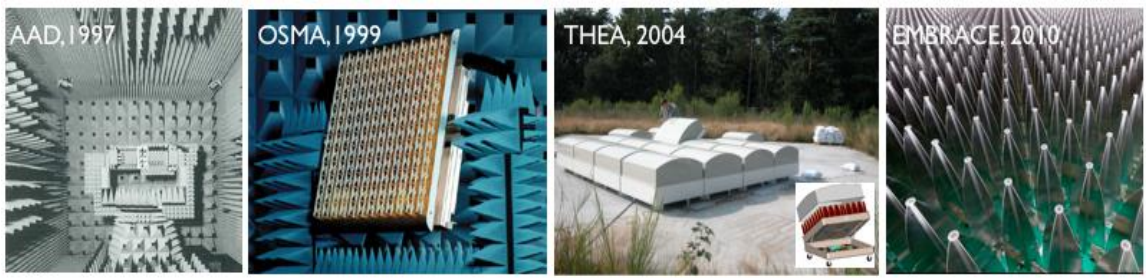

*Fig.1 Some of the engineering test systems that have advanced the development of Aperture Array technology for radio astronomy under the leadership of ASTRON [9].*

Impressive progress has already been made in realising such systems at lower frequencies (typically < 300 MHz) e.g. LOFAR [2], its "all-sky" derivative AARTFAAC [3], the MWA [4] and LWA [5]. At these low frequencies the field of view is naturally large, and the processing (sampling) requirements are such that an all-sky capability is possible with various restrictions (e.g. baseline length, bandwidth, etc.). However, a densely packed aperture array telescope, operating at cm-(GHz) wavelengths (a mid-frequency aperture array) has yet to be realised. An aperture array operating at GHz frequencies (around the so-called "water-hole" 1.4-1.7 GHz) would be very interesting for many science programmes, including SETI. At these frequencies the galactic emission is at a minimum, and there are some who argue this is a "magic" frequency or band since it straddles the HI and OH spectral lines. A significant technical challenge is to make such an array affordable, especially in terms of the power requirements.

Various small-scale engineering systems have shown good results [6], and most recently the EMBRACE system has demonstrated that a large-scale aperture array system capable of making competitive astronomical measurements is now within reach [7, 8]. What is absolutely clear is that so long as data sampling and processing costs continues to fall, the construction of a scientifically productive large aperture array operating at GHz frequencies becomes more of a question of not if, but when. Fig. 1 presents the advance of aperture array technology operating at cm wavelengths through the last 2 decades.

## 2. SETI event rates and the need for field-of-view

Searches for SETI signals have been on-going for many decades [10]. So far these searches have proven unsuccessful e.g. [11], despite the fact that the figure of merit of the associated instrumentation has increased by at least 6 orders of magnitude. In particular, major advances have been made in terms of raw sensitivity, instantaneous bandwidth and spatial, spectral & temporal resolution. These advances, coupled with much greater back-end processing power have permitted radio astronomers to detect a range of periodic or transitory phenomena to be discovered such as pulsars, singular giant pulses, rotating radio transients, and most recently, Fast Radio Bursts (FRBs) [12].

The lack of progress in detecting a SETI signal suggests that emissions from advanced extraterrestrial technologies are extremely rare. Such a conclusion is reinforced by the pristine nature of our own solar system and the lack of evidence for the signatures of Kardashev Type II and Type III civilisations [13] in astronomical data [14] – [17]. All the evidence suggests that advanced civilisations (even Type I civilisations) located in the Milky Way Galaxy might be very rare indeed.

The detection of civilisations similar to our own (Type 0.7) will also be difficult since the radio signals are likely to be relatively faint, even if they are located nearby. If the SETI source counts are similar to cosmic source populations (e.g. AGN), we might expect to find a few very rare events that occasionally lie well above our sensitivity threshold but many more events that lie below it. If this scenario is correct, any reasonable SETI Figure of Merit must strive to maximize both the telescope's sensitivity but also its field-of-view. Compared to other metrics (e.g. instantaneous bandwidth and spectral resolution), field-of-view is one area in which progress has been modest, at least until relatively recently.

## 3. All-sky Radio SETI

Sir Arthur C. Clarke was the first person to consider all-sky radio SETI when he introduced the Argus telescope in his book *Imperial Earth* [18]. Argus would "*look in all directions simultaneously*" and be made up of "*thousands of elements – little more than stiff wires*". Clarke also describes the "*switching and phasing that would allow Argus to swing its antenna-spines electronically – without moving them physically*" with the scale of the instrument being "*at least 1000km in diameter*". His motivation for an "all-sky" radio telescope was similar to our previous discussion - he assumed that SETI signals are rare and perhaps rarer still, if radio waves represent the lowest form of cosmic communication systems, being only a passing phase in the evolution a species technical capabilities.

Argus would work at very long wavelengths (10 - 100s of km), even longer than the SKA precursors and pathfinders mentioned earlier. Clarke realised that in order to detect these kHz radio waves and to achieve any reasonable spatial resolution, Argus would have to be located in deep space (the outer-regions of the solar system - far from the Sun and other highly ionized regions) and be built on an enormous scale.

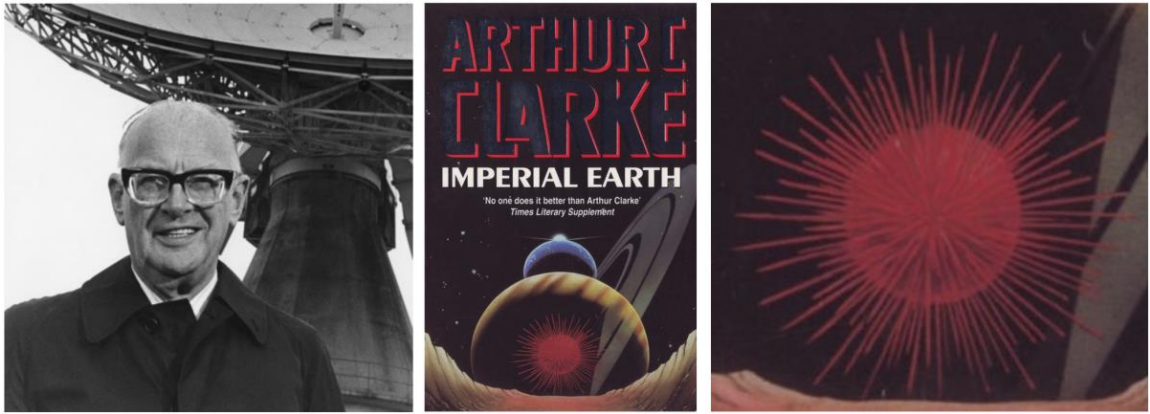

*Fig.2 Sir Arthur C. Clarke (left) and his book "Imperial Earth" (centre). An enlargement of Clarke's all-sky SETI telescope "Argus" is shown (right). Clarke described the core of Argus as "thousands of elements - little more than stiff wires - would radiate from it, like – like the spines of a sea-urchin. Thus it could comb the entire sky for signals".*

## 4. MANTIS

Mid-Frequency Aperture Arrays (MFAA) are expected to form a major component of the next generation technology required to realise the SKA-2 – in particular it's hard to conceive of a conventional dish technology providing a collecting area of 1 square km for any reasonable cost. MANTIS (Mid-Frequency Aperture Array Transient and Intensity-Mapping System) [20] is a proposed SKA-2 precursor, that could represent the first step towards an all-sky radio SETI capability at GHz frequencies. The MANTIS telescope would be located in South Africa, alongside the SKA-1 and the SKA1 precursor, MeerKAT.

The current ambition is to realise a collecting area of around 1500-2000 square metres using of order 250000 antennas. The preliminary specifications of the MANTIS telescope are presented below [20]:

- System Equivalent Flux Density (SEFD): 74 - 44 Jy.
- Frequency range: 0.45-1.45 GHz.
- Bandwidth: 500 MHz.
- Optical Field-of-View: 200 sq. degrees at 1 GHz.

It should be noted that the optical field-of-view is set only by the beam-forming system employed. With current processing limitations, groups of individual antennas have to be analogue beam-formed into "tiles" – for MANTIS the full collecting area might comprise ~ 2500 tiles. The real field-of-view available to an astronomer also depends on the size of field that can be readily

processed – in this regard it is possible to trade bandwidth for field-of-view – currently the goal is to provide a processed field-of-view of ~ 100 square degrees at 1 GHz at full bandwidth.

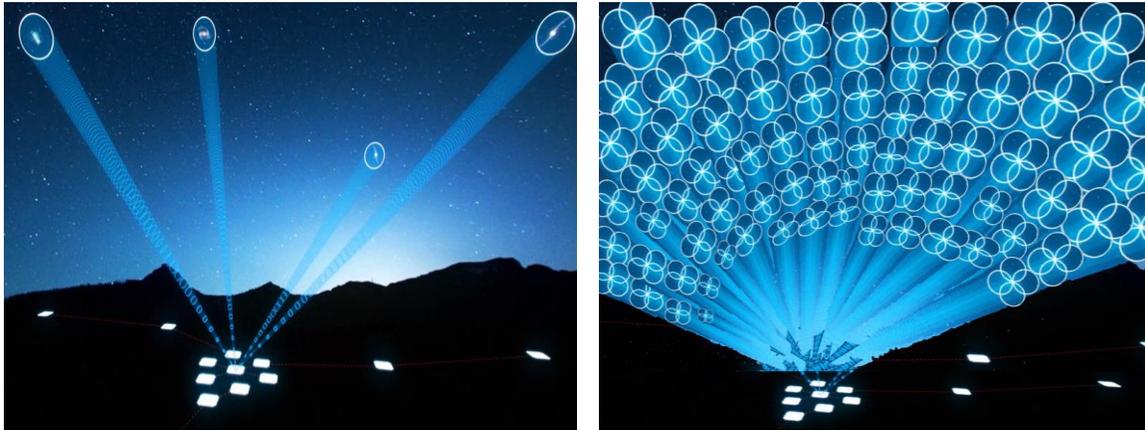

*Fig.3 Given the availability of enough processing power, multi-beam aperture arrays (left) operating at GHz frequencies will eventually provide an "all-sky" capability that could transform SETI research.*

The proposed location of MANTIS is at the SKA site in South Africa, specifically the area known as "K4". This has the advantage that the telescope can also act as an "early-warning" transient detector for both MeerKAT and SKA1-mid, similar to the role played by MeerLICHT in the optical domain (see Groot et al. these proceedings). Apart from transient and pulsar science, one of MANTIS' main drivers is HI Intensity Mapping. SETI is also a stated goal of the science programme with the large field-of-view seen as a first step towards an all-sky radio SETI capability at these frequencies.

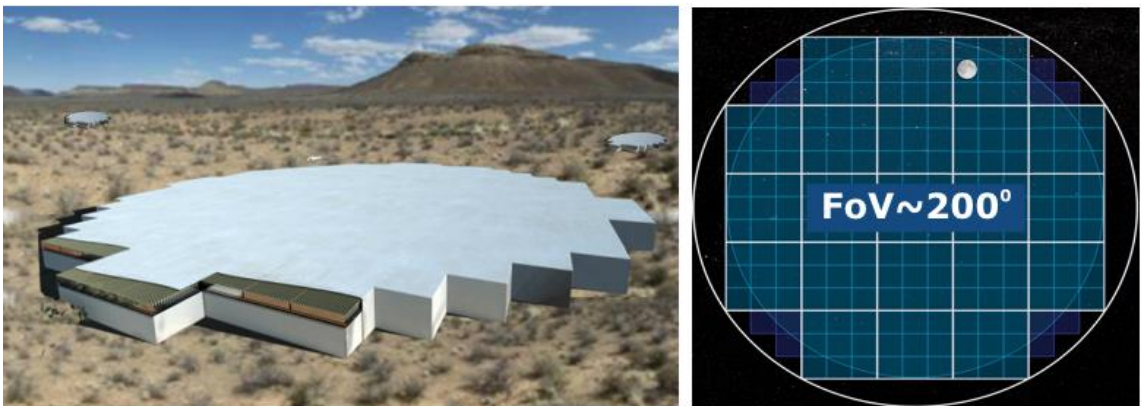

*Fig.4 MANTIS located at the South African SKA site "K4" (left). The optical field-of-view of this telescope is presented with the Moon presented for scale (right).*

For the most part, SETI observations performed by MANTIS could run commensally with other astronomical observations. Commensal SETI surveys would require a data spigot to be in place that would provide a SETI backend with copies of the various beam-formed data – a separate SETI backend is required in order to meet the extreme frequency (and time) resolution requirements of narrow-band signal searches (typically of order 1 Hz). In addition to this commensal approach, SETI researchers would also require modest amounts of targeted observing time for objects of special interest (e.g. the conjunction of extrasolar planetary systems along the Earth's line-of-sight [21]).

## 5.    Conclusions

The emergence of highly capable and flexible Mid-Frequency Aperture Arrays (MFAA) with a large field-of-view can be very important for all manner of astronomical research, but in particular SETI. The proposed MANTIS precursor represents a step towards realising an all-sky radio SETI capability. If SETI events are both rare and transient in nature, an all-sky instrument could represent the type of technology breakthrough that transforms the field. The realisation of Sir Arthur C. Clarke's vision of an all-sky SETI radio instrument is simply a matter of time.